# Surface Modification of Mild Steel by Carbon Nanotubes Reinforcement under Electron Beam Melting

Munawar Iqbal, Adeela Nairan and M. Ayub Faridi

Centre for High Energy Physics, University of the Punjab, Quaid-e-Azam Campus, Lahore 54590, Pakistan

**Abstract:**

Structural properties of mild steel (MS) have been modified by embedding Carbon Nanotubes (CNT's) in its matrix. The embedment was done by forming a composite at the surface of MS with the help of a thermionic electron beam source at an acceleration potential of 5 kV and beam current of 45 mA. Structural analysis was carried out by X-Ray Diffraction (XRD). Stress and strain were reduced about three times while the crystalline size was reduced to 22% after embedment of CNT's. Surface morphology was observed by Scanning Electron Microscope (SEM) and composition was examined by Energy Dispersive X-Ray Spectroscopy (EDX) respectively. Average grain size of the modified sample was reduced from 1.94 μm to 720 nm whereas the hardness was increased to 30% for as received samples.

# 1. Introduction

Mild steel is extensively being utilized in various applications due to its strength, low cost, ductility and weld-ability [1]. There are many industrial applications which require high strength such as in car bodies, ships, fittings of overhead electric transmission lines and greatly in construction [2]. Nevertheless, low hardness confines its applications but, this deficiency could be overcome by applying surface modification techniques [3-8]. Wide research has been made over surface modification by formation of composite with mild steel. Munawar et. al. [9-10] has studied the mechanical properties of mild steel reinforced with SiC and $B_4C$ under electron beam melting. Their findings have shown increase in hardness and variation in lattice parameters. Majumdar [11] has made in-situ Iron Silicide coating on mild steel using $CO_2$ laser re-melting. He found remarkable increase in hardness and wear resistance of the material. Jongmin et. al., [12] have made a composite of stainless steel with TiC particles irradiated by electron beam. They also examined hardness and wear resistance of stainless steel. Similarly, Chandra, et.al., [13] have modified mild steel surface with SiC using two step laser cladding with $CO_2$ laser. They have shown increase in hardness and also significant increase in wear and corrosion resistance. The purpose of these modifications was to enhance the mechanical properties of MS to make it at par to costly materials like aluminum, titanium, etc.

However, in literature, no description is accessible on surface modification of MS reinforced with carbon nanotubes (CNT's) under electron beam melting. CNT's have extremely high Young's modulus, large fracture and elastic strains sustaining ability. The theoretical modulus of elasticity of single wall CNT's was assessed approximately 5 TPa. CNTs are supposed to be perfect for developing unconventional and extraordinary performance composites. CNT's reinforcement increases the mechanical properties, as shear strength [14], tensile strength [15], flexural strength, [16] fatigue resistance while reduces the crack growth rate [17] and interlaminate fracture resistance.

In our present work, effort has been made to modify MS surface, reinforced with CNT's by the application of a thermionic electron beam. The aim of the study is to modify the mild steel surface and change its mechanical properties keeping the bulk of the material unaffected. Here, we have used single walled CNT's for reinforcement. These nanotubes were considered as ideal because they possess very small diameter (below 1 nm), high tensile strength and Young's modulus as compared to multi walled CNT's.

## 2. Experimental

Mild steel having code 8620 was used as a target for mechanical and structural investigations. Sample with dimensions (12 x 12 x 6 mm$^3$) were cut down from mild steel sheet. It was cleaned after cutting through grinding and then polished with diamond paste upto 1 µm. In order to embed CNT's, grooves of 1x 1.5 mm$^2$ were made on the surface of MS. Then CNT's were added to the grooves by making slurry with hexane at the surface of MS samples. Finally, samples were placed beneath the electron gun source for melting. A continuous thermionic electron beam was allowed to fall normal to the surface with operating current of 45 mA, 5 kV of acceleration potential, 2 MHz frequency and vacuum pressure of 1E$^{-4}$ mbars.

A schematic diagram of the experiment is given in Fig. 1. During whole process, electron beam chamber was kept under vacuum (10$^{-4}$ mbar) and in contact with chiller to have high cooling rate. To obtain fine microstructure, the reinforced sample was again polished and etched with 3% Nital solution for 30 sec. For microstructure analysis; X Ray Diffraction (XRD) and Scanning Electron Microscope (SEM) were carried out. Chemical composition was examined by Energy Dispersive X-ray Spectroscopy (EDX) and Vickers hardness tester was used to measure hardness of the modified sample.

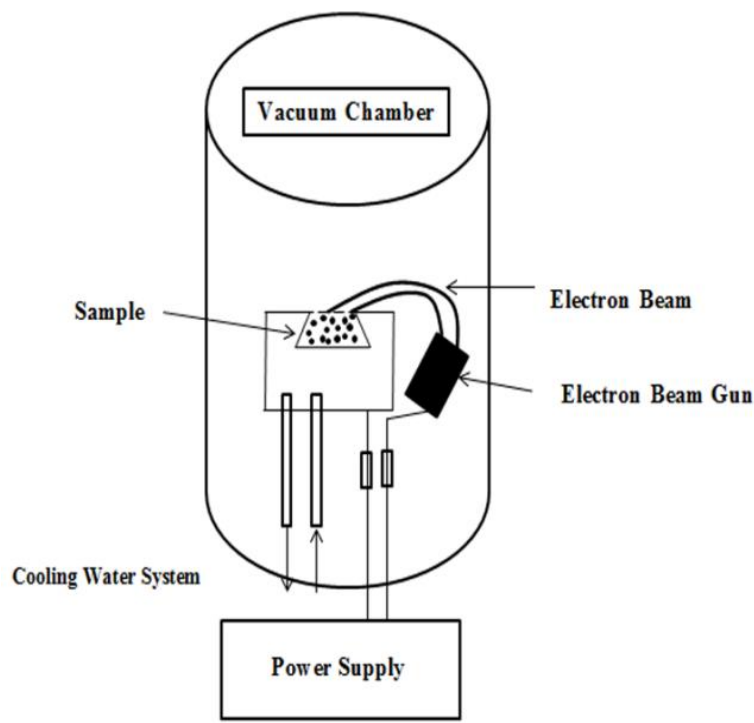

**Fig. 1.** *Schematic diagram of the experimental setup*

## 3. Results and Discussion

XRD pattern of mild steel before and after reinforcement of CNT's are shown in Fig. 2. It can be seen that the significant peaks of mild steel before and after are from iron (110). Other dominant peaks are of FeCr (200) and FeCr (211). Peak of iron before and after modification is 100% intensive. First and second peak of iron chromate before embedment are 4 and 11% intensive while after embedment are 11 and 17% intensive respectively. XRD data taken from JCPDS cards for the mild steel before and after reinforcement of CNT's is shown in Table 1.

**Table 1:** XRD data before and after embedment of CNT

| Serial No. | 2-Theta | JCPDS Card 2-Theta | d-values | Phases Identified |
|---|---|---|---|---|
| 1. | 44.669 | 44.677 | 2.02873 | Fe |
| 2. | 65.062 | 64.957 | 1.43363 | FeCr |
| 3. | 82.376 | 82.244 | 1.16973 | FeCr |

To calculate crystalline size and lattice constant, X-ray diffraction (XRD) analysis were done. Crystalline size of as received mild steel was calculated through Scherer formula and the measured value was t= 435.7 $^oA$. After the embedment of CNTs, it was calculated to be t=340.95 $^oA$ . Crystal structure of iron was cubic and it was observed that the calculated value for lattice constant was 2.872 $^oA$. The crystal structure remained cubic and measured value for lattice constant was 2.867 $^oA$ which is approximately near to the standard value 2.8664 $^oA$.

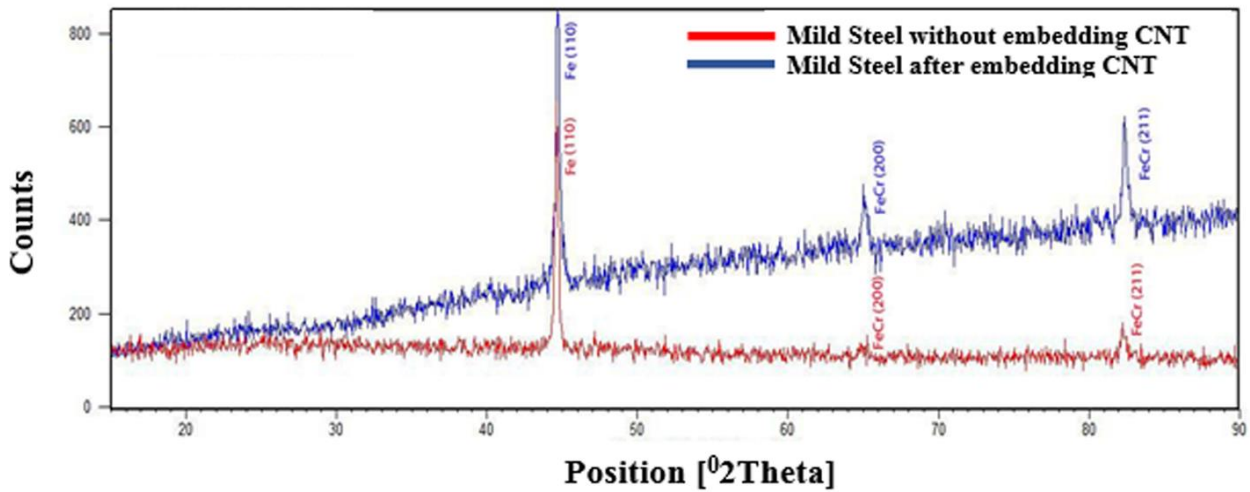

**Fig. 2.** XRD of mild steel before and after embedment of CNT

The measured values for the lattice constants of mild steel before and after the reinforcement of CNTs verified the investigated peaks.

Stress and strain are important parameters to understand behavior of the crystal. Strain calculated by formula

$$\varepsilon = \frac{d_n - d_0}{d_0}$$

and it was 0.003 and 9.87E$^{-4}$ before and after embedment of CNTs respectively. $d_n$ and $d_o$ represents standard and calculated values which were 2.0268 and 2.0328 in case of as received and were 2.026 and 2.028 after the embedment, respectively. Stresses were calculated by using Young's modulus which was 210 GPa for MS. Therefore, value of the stress was calculated by using relation " Stress = Strain x Young's Modulus" (taking values of strain and Young's modulus from the above measurements), and it was 0.622 and 0.207 GPa, respectively.

It was concluded that, intensity of the peaks of mild steel increased after the reinforcement of CNT's on its surface. Intensity of peaks raised due to several reasons; first is the embedment of CNT's and secondly by increasing temperature through the high density electron beam that has the capability of localized heating as well. Due to CNT's phase shift is produced in all the peaks. Crystalline size has been reduced to 21.72 % and lattice constants are approximately equal to the standard values. Strain and stresses were reduced approximately 3 times after embedment of CNT's. These modifications are reported in Table 2. For hardness measurements, a weight of 10 kg was applied for 10 sec using Vickers hardness method. The measured value of hardness before and after the embedment of CNTs was 145 and 204 HV respectively. A hardness comparison is shown in Fig. 3. The hardness of the material has been increased about 30% compared to as received sample. To investigate surface morphology and grain size of mild steel, images were taken from scanning electron microscope (SEM). SEM micrograph of as received sample is shown in Fig. 4. The grain size was calculated by designing a particular geometry in selected region and was approximately 1.94 μm. EDX analysis of the sample was performed in the same selected region through which the image of SEM was taken. This spectrum captured at particular region is shown in Fig. 5.

It has been observed that different composition of elements is present in the small region of grain boundary. Iron is a major element in two intensive peaks whereas chromium and cobalt are present as a shoulder peaks of iron. Cobalt and oxygen are also present but in small ratio whereas Calcium is absent. The atomic percentage of cobalt, chromium, oxygen and iron are 3.76, 11.33, 19.05, and 65.85 respectively.

**Table 2: Comparison of structural and morphological parameters**

| Serial No. | Properties | Values before embedment | Values after embedment | % variation |
|---|---|---|---|---|
| 1 | Crystalline Size | 435.7 Å | 340.95 Å | 21.7% decrease |
| 2 | Lattice Constant | 2.872 Å | 2.867 Å | 1.74% decrease |
| 3 | Stress | 0.622GPa | 0.207GPa | 66.7% decrease |
| 4 | Strain | 0.003 | 0.000987 | 67.1% decrease |
| 5 | Grain Size | 1.94μm | 720nm | 62.8% decrease |
| 6 | Hardness | 145HV | 204HV | 28.9% Increase |

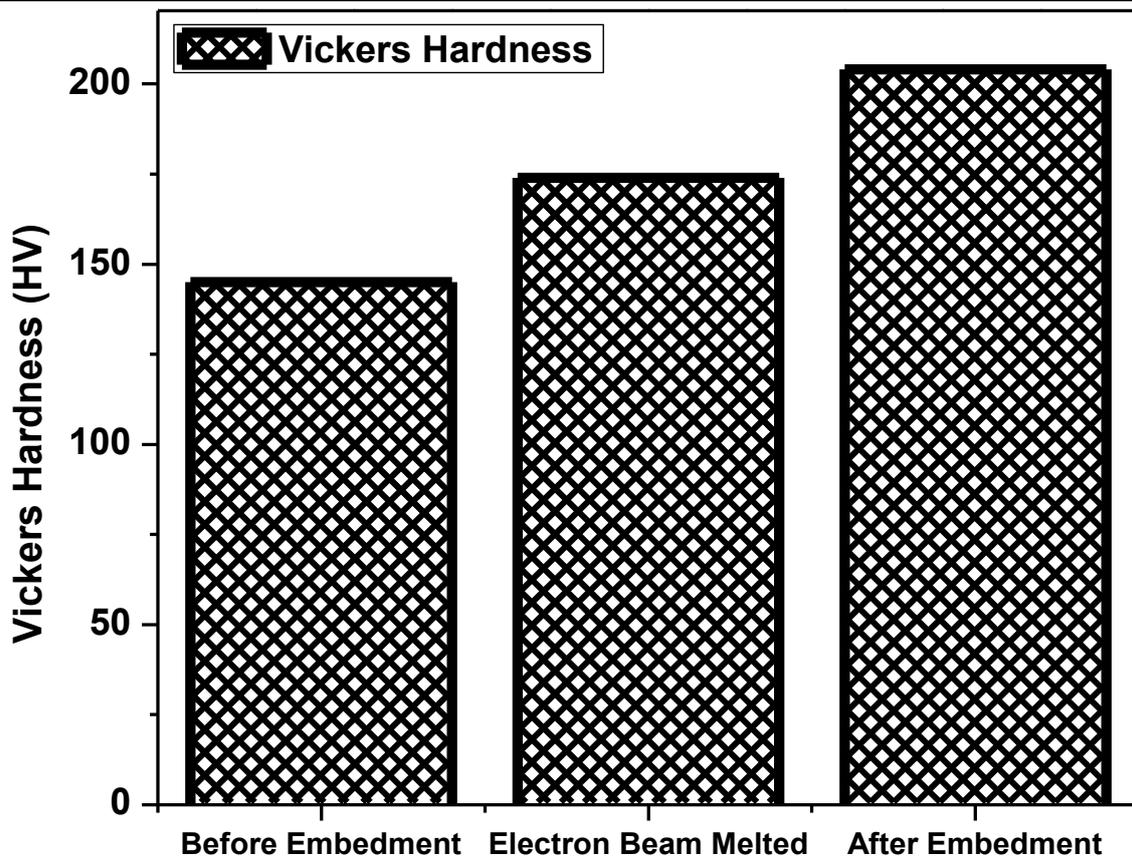

**Fig. 3.** Vickers hardness comparison

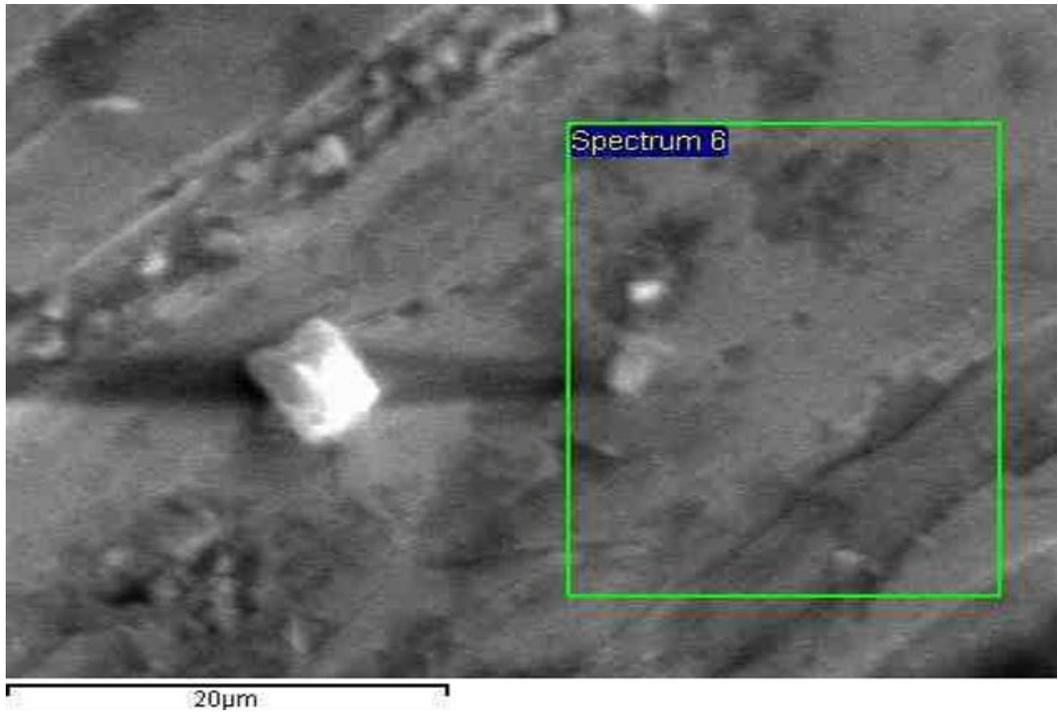
**Fig. 4.** SEM Image of mild steel without embedment of CNT

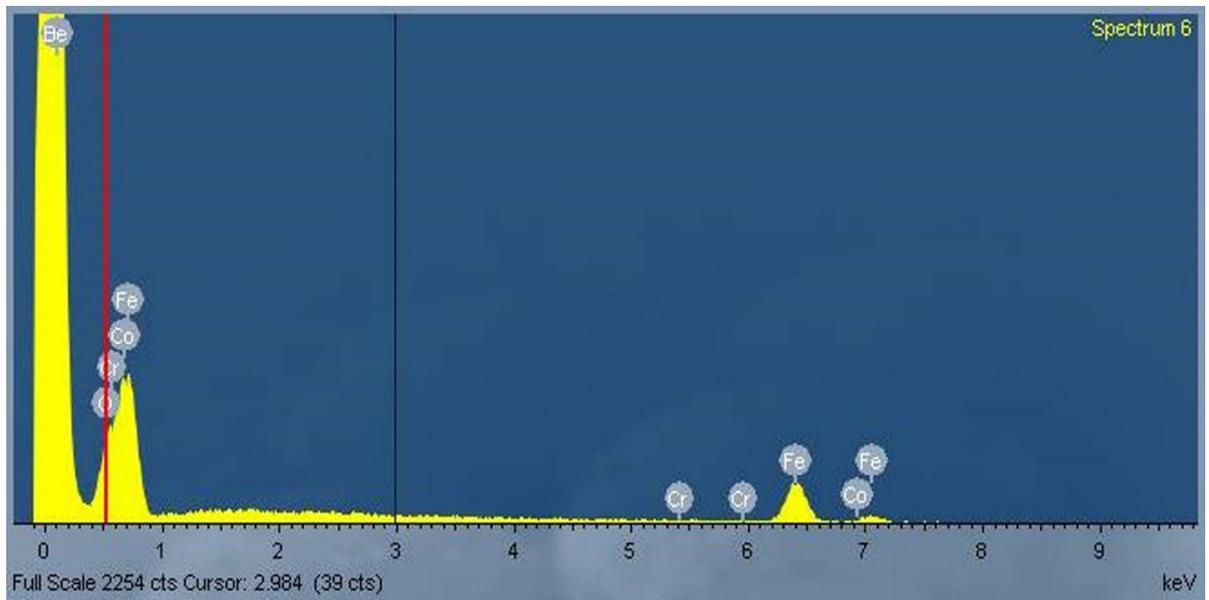
**Fig. 5.** *Spectrum taken from EDX*

Surface morphology and grain size after reinforcement was again calculated by SEM. This time accelerating potential was increased from 10 kV to 25 kV as shown in Fig. 6(a). This image confirmed the presence of CNT's which were observed clearly in Fig.6 (b). The calculated grain size of the selected CNT's was approximately 450 nm and 990 nm respectively. Therefore, the average grain size after embedment of CNT's was 720 nm.

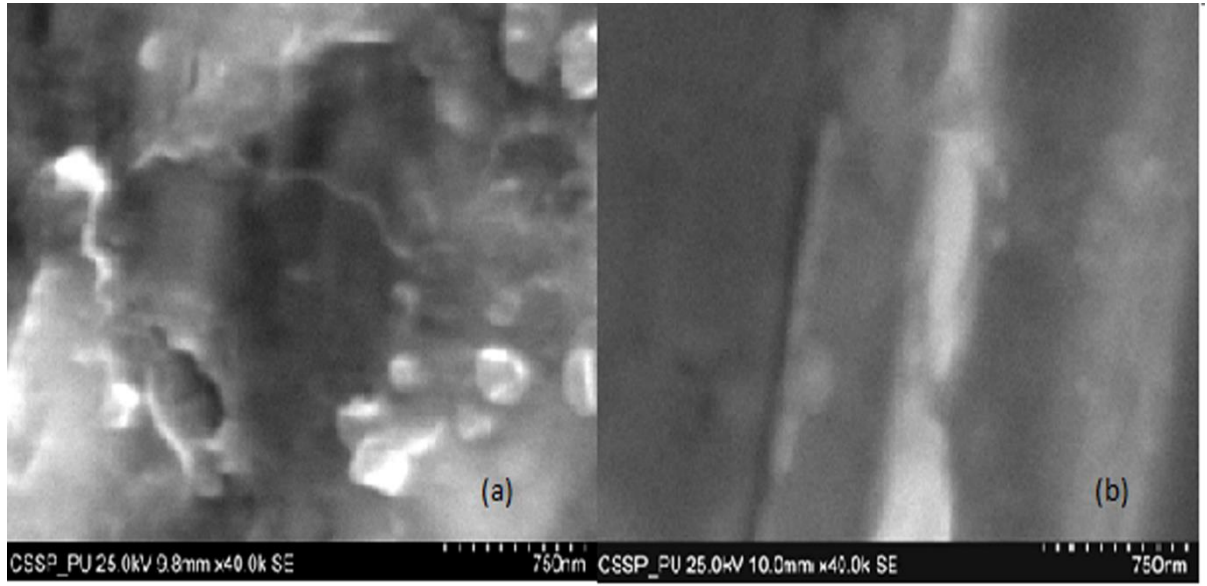

**Fig. 6.** SEM Image of mild steel after embedment of CNT

## 4. Conclusions

We have successfully modified the crystal structure of MS by adding CNT's using electron beam melting by forming a composite at its surface. SEM confirmed the existence of CNT's in the matrix of the MS. As a result, crystalline size of the sharp peak of iron has been reduced to 21.72 %; average grain size was reduced from 1.94 μm to 720 nm. Also stress and strain were reduced approximately 3 times compared to as received sample. The measured lattice constant before and after the reinforcement of CNT's were 2.872 $^oA$ and 2.867 $^oA$.. These values are approximately equal to the standard data. Hardness of mild steel was increased about 30% after the reinforcement of CNT's. Hence, electron beam melting is a high-quality mechanism to change the mechanical properties of MS due to its localized heating capability. Moreover, electron is the lightest particle and its energy can be controlled by the applied acceleration potential. So it gives huge amount of energy without dissipation. As mild steel is intensively being used for industrial applications, hopefully this work will open new research aims to manufacture mild steel products with enhanced microstructure and mechanical properties at optimized cost.

**Acknowledgments:**

We are very grateful to the Physics laboratory, Lahore College for Women University, Pakistan; Centre for excellence of Solid State Physics, Punjab University, Lahore, Pakistan; Islamia University, Bahawalpur, Pakistan and Korea Institute of Materials Sciences, Republic of Korea.